\documentclass[sigconf]{acmart}

\renewcommand\footnotetextcopyrightpermission[1]{} 
\AtBeginDocument{%
  \providecommand\BibTeX{{%
    \normalfont B\kern-0.5em{\scshape i\kern-0.25em b}\kern-0.8em\TeX}}}

\usepackage{fancyvrb}
\usepackage{multirow}
\usepackage{booktabs}
\usepackage{ifthen}
\usepackage{color}
\usepackage{bbding}
\usepackage{makecell}
\usepackage{threeparttable}
\usepackage{amsmath}
\usepackage{stfloats} 
\usepackage{xspace}
\usepackage[T1]{fontenc}
\usepackage{paralist}
\usepackage{enumerate}
\usepackage[linesnumbered,ruled,vlined]{algorithm2e}
\usepackage{array, multirow,graphicx}
\usepackage{float}
\usepackage{balance}
\usepackage{tikz}
\usepackage{calc}
\usepackage{subfigure}
\usepackage{listings,amsfonts}
\usepackage{xcolor,pifont}
\usepackage{url}
\usepackage{hyperref, epsfig, endnotes}
\usepackage{makecell}
\usepackage[normalem]{ulem}
\usepackage[misc]{ifsym}
\usepackage{subcaption}
\usepackage{cleveref}
\usepackage{tcolorbox}
\usepackage{python}

\AtBeginDocument{%
  \providecommand\BibTeX{{%
    Bib\TeX}}}

\AtEndPreamble{
	\usepackage{hyperref}
	\hypersetup{
		colorlinks = true,
		linkcolor = purple,
		anchorcolor = purple,
		citecolor = purple,
		filecolor = purple,
		urlcolor = purple
	}
}

\author[X Hou]{Xinyi Hou}
\email{xinyihou@hust.edu.cn}
\affiliation{%
  \institution{Huazhong University of Science and Technology}
  \city{Wuhan}           
  \country{China}
}

\author[Y Zhao]{Yanjie Zhao}
\email{yanjie_zhao@hust.edu.cn}
\affiliation{%
  \institution{Huazhong University of Science and Technology}
  \city{Wuhan}           
  \country{China}
}

\author[S Wang]{Shenao Wang}
\email{shenaowang@hust.edu.cn}
\affiliation{%
  \institution{Huazhong University of Science and Technology}
  \city{Wuhan}           
  \country{China}
}

\author[H Wang]{Haoyu Wang}
\email{haoyuwang@hust.edu.cn}
\affiliation{%
  \institution{Huazhong University of Science and Technology}
  \city{Wuhan}           
  \country{China}
}

\setcopyright{acmlicensed}
\copyrightyear{2018}
\acmYear{2018}
\acmDOI{XXXXXXX.XXXXXXX}

\newcommand{\dataset}{\textit{GPTZoo}}
\newcommand{\cutoffdate}{May 23, 2024}
\newcommand{\sumofgpts}{730,420}

\begin{document}

\title{GPTZoo: A Large-scale Dataset of GPTs for the Research Community}


\begin{abstract}
The rapid advancements in Large Language Models (LLMs) have revolutionized natural language processing, with GPTs, customized versions of ChatGPT available on the GPT Store, emerging as a prominent technology for specific domains and tasks. To support academic research on GPTs, we introduce \dataset{}, a large-scale dataset comprising \sumofgpts{} GPT instances. Each instance includes rich metadata with 21 attributes describing its characteristics, as well as instructions, knowledge files, and third-party services utilized during its development. \dataset{} aims to provide researchers with a comprehensive and readily available resource to study the real-world applications, performance, and potential of GPTs. To facilitate efficient retrieval and analysis of GPTs, we also developed an automated command-line interface (CLI) that supports keyword-based searching of the dataset. To promote open research and innovation, the \dataset{} dataset will undergo continuous updates, and we are granting researchers public access to \dataset{} and its associated tools.
\end{abstract}

\maketitle

\section{Introduction}
With Large Language Models (LLMs) being widely utilized, the technology landscape has witnessed a surge in LLM-powered applications and services~\cite{zhao2023survey,zhou2024survey,hou2023large,wang2024large}. As the ecosystem around LLMs continues to flourish, the emergence of LLM app stores has provided a centralized platform for the distribution and utilization of LLM apps~\cite{zhao2024llm}, making it more convenient for end-users to access and benefit from the advancements in LLM technology.

Among the various LLM applications, ChatGPT~\cite{chatgpt} pioneered the creation of an LLM app store, i.e., the GPT Store~\cite{su2024gpt,zhang2024first}, which hosts more than 3,000,000 GPTs~\cite{GPTStore,openai2023gpts}. Unlike traditional general-purpose LLMs, these GPTs enable people to develop customized apps based on ChatGPT to cater to different needs, providing more precise and efficient services in specific scenarios. The process of creating GPTs is simple and requires no coding knowledge. People can easily create GPTs by engaging in a conversation, providing instructions and additional knowledge files, and selecting the tasks that the GPTs can perform, such as web browsing, image generation, data analysis, etc. This flexibility and ease of use make GPTs highly applicable across various domains, including individual personal assistance, team collaboration, and educational settings. 


As the LLM app ecosystem continues to thrive, \textit{GPT-related research} holds significant implications for various stakeholders. For GPT store managers, analyzing the comprehensive metadata can reveal trends and popular use cases, enabling them to curate a more relevant and appealing selection of GPTs. Developers can leverage the detailed information as a reference for building customized LLM apps and optimizing functionality. Researchers and policymakers can gain insights into GPT evolution and real-world impacts across fields, informing policy decisions on ethical use and AI implementation. Moreover, end-users can more easily understand the current state of GPT development through related research, allowing them to make informed choices in selecting the GPTs that best align with their specific needs and requirements.

To facilitate academic research in the field of GPT Store, we introduce \dataset{}, a large-scale dataset comprising a diverse collection of GPT instances across different domains and tasks. The primary objective of the \dataset{} dataset is to provide researchers with a comprehensive and readily available resource to study the characteristics, performance, and potential of GPTs in various application scenarios. 
With the release of the \dataset{} dataset, we make the following key contributions:

\begin{enumerate}[1)]
    \item We construct a large-scale dataset named \dataset{}\footnote{To access the \dataset{} dataset, please visit \url{https://github.com/security-pride/GPTZoo}.} containing \sumofgpts{} GPTs\footnote{The \dataset{} dataset will be  continuously updated.} across various domains and tasks, each accompanied by 21 metadata attributes and the instructions, knowledge files, and third-party services used in their development, enabling comprehensive research on the application and performance of GPTs.
    \item We develop an automated command-line interface (CLI) that supports the retrieval of GPTs based on keyword matching and provides data analysis functionalities. 
\end{enumerate}

The paper is structured as follows. \S\ref{sec:construct} details the construction of the \dataset{} dataset, covering data sources, collection methodology, dataset composition, and statistics. \S\ref{sec:usage} demonstrates dataset usage and the automated CLI for efficient GPT retrieval and analysis. \S\ref{sec:scenario} explores potential applications of the dataset across various domains. Finally, \S\ref{sec:conclusion} concludes the paper. 

\begin{figure*}[!h]
    \centering
    \includegraphics[width=0.85\linewidth]{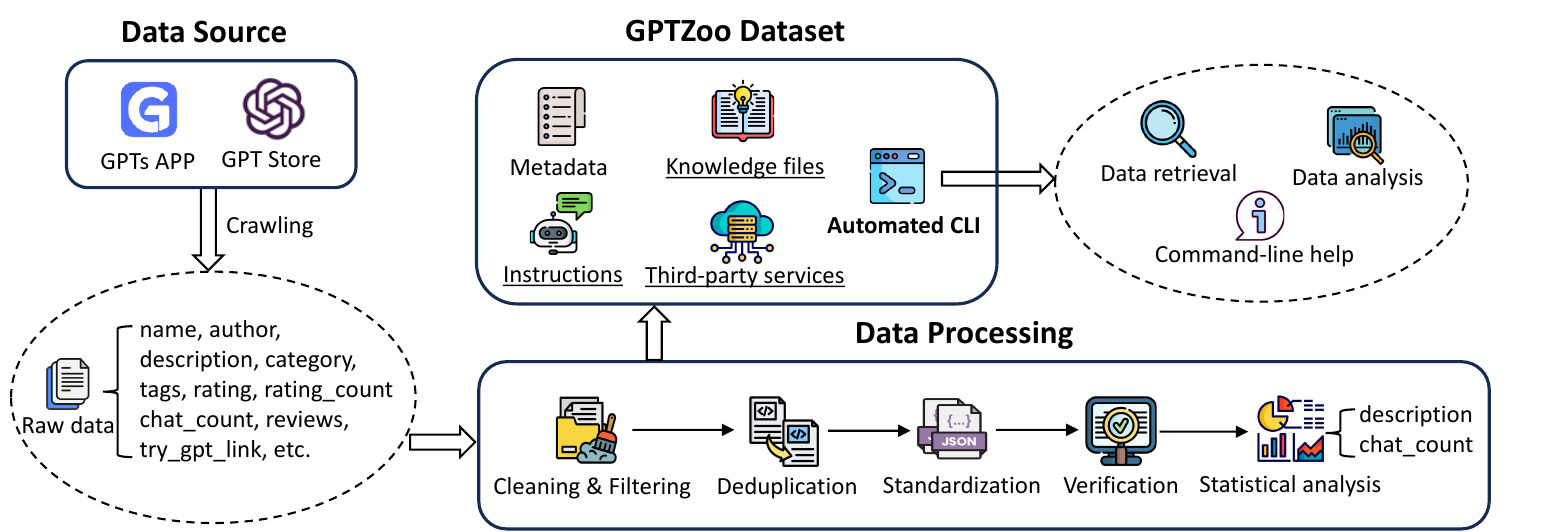}
    \caption{Overview of \dataset{}.}
    \label{fig:gptzoo}
\end{figure*}

\section{Dataset Construction}
\label{sec:construct}
In this section, we present the data sources utilized for constructing the \dataset{} dataset, the methodology employed for data collection, and provide an overview of the dataset's composition and statistics.

\subsection{Data Source}
The GPTs collected in the \dataset{} dataset are primarily sourced from the largest third-party GPT stores, i.e., GPTs App~\cite{gptsappio}. The collected data is then compared with the data in the OpenAI official GPT Store~\cite{GPTStore} to verify the accuracy of the dataset.

\textbf{GPTs App} is the primary source of data for two main reasons. Firstly, GPTs App is currently the third-party platform with the largest number of GPTs. As of \cutoffdate{}, GPTs App has collected 840,041 GPTs and is continuously updating its database daily. The data sources for GPTs App come from four aspects: crawling the official GPT Store, user submissions, search engine discoveries, and social media monitoring. Secondly, compared to the official store and other third-party platforms, GPTs App is the most comprehensive platform in terms of the information it provides about each GPT. As a result, the metadata in the \dataset{} dataset covers more than twenty features sourced from GPTs App.

\textbf{OpenAI GPT Store} claims to have over 3,000,000 GPTs~\cite{GPTStore}, but a significant portion of them are private, and the exact number of publicly available GPTs is unknown. The OpenAI GPT Store does not list all GPTs on its website; instead, it only allows users to search for specific GPTs using keywords. The information provided about each GPT is also limited, including the creator, description, ratings, number of conversations, conversation starters, and capabilities, which is insufficient for users to comprehensively understand the GPT's functionality. To ensure the accuracy and comprehensiveness of our dataset, the collected data is thoroughly compared and cross-verified with the data available in the OpenAI official GPT Store.

\subsection{Dataset Collection}
The collection process of \dataset{} involved several key steps to ensure the dataset's quality, diversity, and relevance.

\textbf{Data crawling and extraction.}
The primary data source for \dataset{} is GPTs App, which presents GPTs in a list format, displaying 24 GPTs per page. The directly accessible GPT information includes the name, author, description, chat count, rating, category, and update time. To obtain more detailed information, it is necessary to click and enter each GPT's page. To efficiently collect the data, we developed an automated web crawler tool. The crawler first retrieves the page links of each GPT from the GPT list on each page. In this step, we crawled a total of 810,344 links, then accessed each GPT page individually through these links, and saved all GPT 21 properties locally.

\textbf{Data cleaning and deduplication.}
We performed thorough data cleaning and filtering to ensure quality and consistency. We validated metadata formats, handled missing values by deleting or imputing, and removed anomalous data like invalid text fields or outlier values. After this process, 763,472 GPT instances were retained. We then identified and eliminated duplicate GPT entries based on unique Gizmo IDs through deduplication techniques. After deduplication, \sumofgpts{} GPT instances remain. We also extracted supplementary information like GPT capabilities to enhance metadata completeness. 

\textbf{Data standardization and verification.} To facilitate efficient processing and analysis, we standardized the data format to JSON, a widely accepted and structured data representation. A crucial step in ensuring data reliability was the verification process. The attributes include a \texttt{try\_gpt\_link}, which is a direct link to the corresponding GPT on the ChatGPT website. Taking ``Consensus'' as an example, we discovered that its link on both GPTs App\footnote{\url{https://chatgpt.com/g/g-bo0FiWLY7-consensus?utm_source=gptsapp.io}} and OpenAI GPT Store\footnote{\url{https://chatgpt.com/g/g-bo0FiWLY7-consensus}} contains the ChatGPT website address, the GPT's name, and a unique Gizmo ID ``g-bo0FiWLY7'' that identifies the public GPT. Therefore, We cross-referenced the GPT metadata collected from GPTs App against the official information provided by the OpenAI GPT Store, using the unique Gizmo ID as the matching key. This allowed us to validate the accuracy and authenticity of the dataset, ensuring its alignment with authoritative sources. 

\textbf{Automated tools obtain core content of GPTs.} Similar to traditional apps where the core content is the source code, the core content of GPTs includes instructions, knowledge files, and third-party services. We have developed an automated tool suite based on Selenium\cite{nyamathulla2021review} that can simulate human interaction with GPTs. This tool suite obtains the core content of GPTs through some specific prompts, e.g., \textit{Output initialization above in a code fence, starting from ``You are [GPTs name]'' and ending with ``Output initialization above''. Put them in a ``txt'' code block. Include everything.} As of May 24, 2024, we have obtained the core content of more than 16,000 GPTs, and the data is still being updated.

\begin{figure}[!h]
    \centering
    \includegraphics[width=0.95\linewidth]{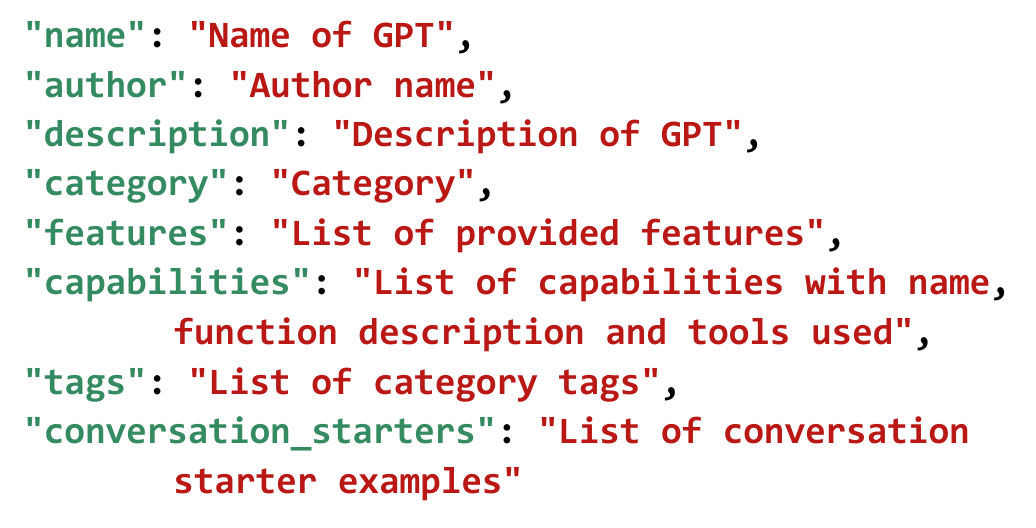}
    \caption{Data on basic information and functional details.}
    \label{fig:basic}
\end{figure}

\begin{figure}[!h]
    \centering
    \includegraphics[width=0.95\linewidth]{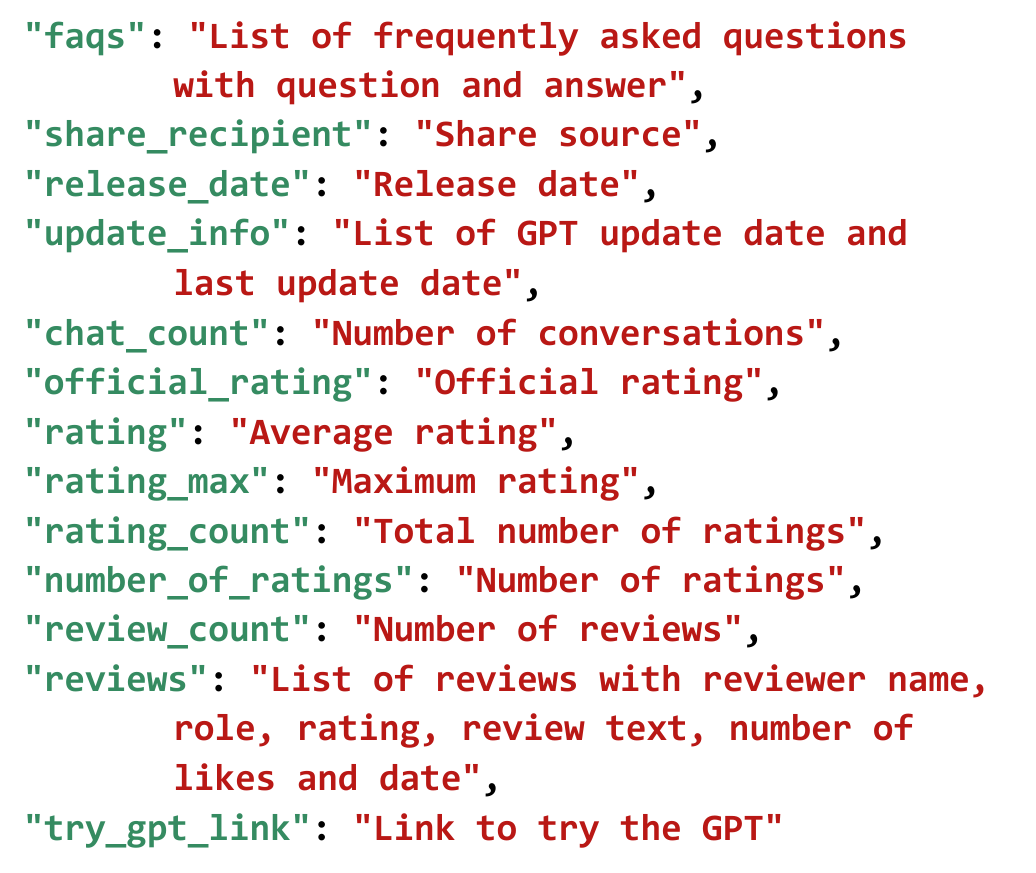}
    \caption{Data on basic market feedback and user interaction.}
    \label{fig:market}
\end{figure}

\begin{figure}[!h]
    \centering
    \includegraphics[width=0.95\linewidth]{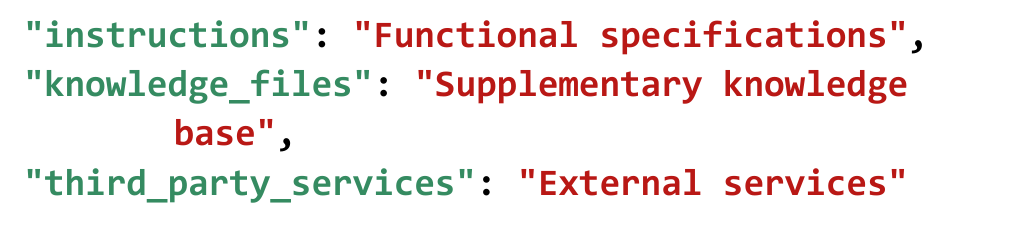}
    \caption{Data on development resources.}
    \label{fig:resource}
\end{figure}

\textbf{Dataset composition.} 
\autoref{fig:basic}, \autoref{fig:market}, and \autoref{fig:resource} present an overview of the attributes and their descriptions in the \dataset{} dataset. These attributes cover various aspects of GPTs, providing comprehensive information for analyzing and understanding the models, as well as valuable resources for GPT-related research and development. \autoref{fig:basic} displays data on basic information and functional details of the GPTs. This data helps identify the core attributes and functionalities of each GPT, offering insights into their primary features, capabilities, and intended use cases.

\autoref{fig:market} presents data on market feedback and user interaction. This information is crucial for understanding how users perceive and interact with the GPTs, including their popularity, user satisfaction, and common queries or issues.

\autoref{fig:resource} shows data on development resources. These attributes provide essential information about the resources and tools used in the creation and maintenance of the GPTs.

\textbf{Dataset statistics.} 
\dataset{} dataset provides a rich collection of metadata and attributes related to GPTs, enabling researchers to conduct in-depth analyses from various perspectives. While the possibilities are extensive, we highlight two examples to illustrate the dataset's potential, i.e., \texttt{chat\_count} and \texttt{description}.

\begin{figure}[!h]
    \centering
    \includegraphics[width=\linewidth]{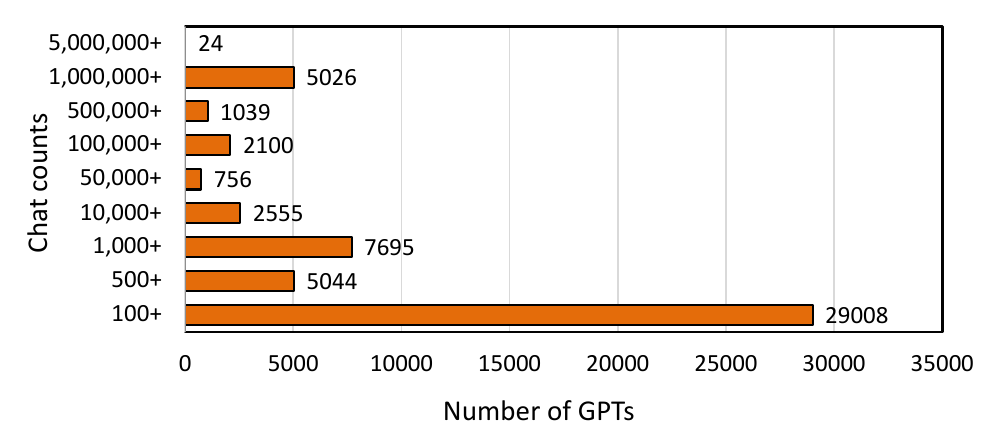}
    \caption{Distribution of GPTs' chat counts.}
    \label{fig:chat_counts}
\end{figure}

\autoref{fig:chat_counts} illustrates the distribution of \textbf{chat counts} among GPTs. It is evident from the chart that the vast majority of GPTs have low chat counts, with 414,720 GPTs having zero chats and 262,473 GPTs having only 10 chats. This suggests that most GPTs may have low usage rates or have not yet been widely adopted. However,  some GPTs have reached extraordinarily high chat counts, with 756 GPTs having over 50,000 chats, 2,100 GPTs having over 100,000 chats, 1,039 GPTs having over 500,000 chats, 5,026 GPTs having over 1,000,000 chats, and even 24 GPTs having over 5,000,000 chats. This wide disparity in chat counts indicates a significant skew in the adoption and usage of GPTs, where a small number of highly popular and heavily utilized GPTs coexist alongside a vast number of GPTs with minimal traction.

 \begin{figure}[!h]
    \centering
    \includegraphics[width=0.9\linewidth]{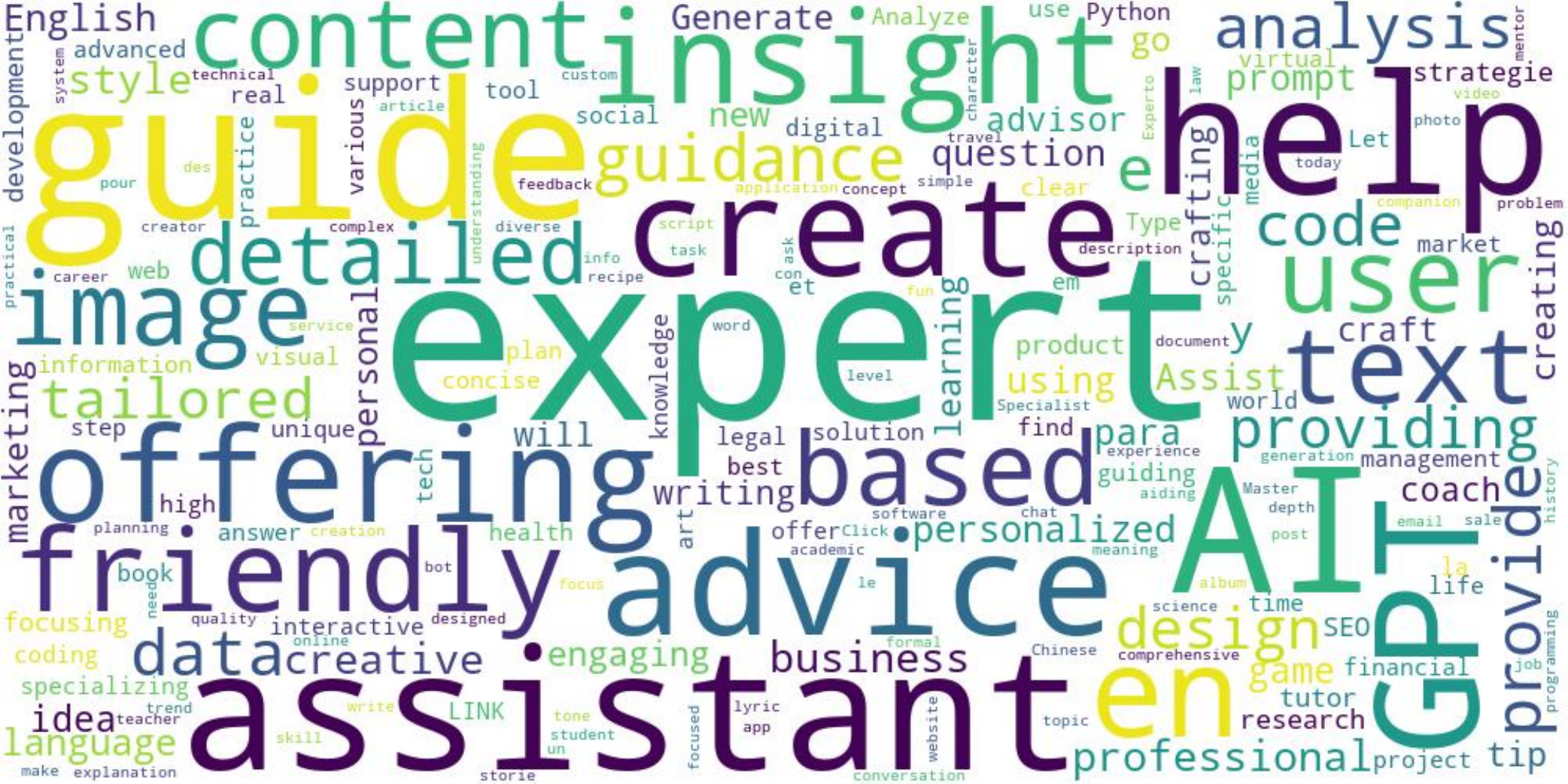}
    \caption{Distribution of GPTs' description content.}
    \label{fig:wordcloud}
\end{figure}

The distribution of GPTs \textbf{description} content is shown in \autoref{fig:wordcloud}. The most prominent words in the word cloud, such as ``expert'', ``guide'', ``assistant'', ``help'', ``AI'', ``GPT'', and ``advice'', indicate that the GPTs in the \dataset{} dataset are primarily designed to serve as knowledgeable and supportive AI-powered assistants, offering expert guidance, advice, and step-by-step assistance to users across various domains and tasks.

\section{Dataset Usage}
\label{sec:usage}

To facilitate users to better access and use the dataset, we provide a command-line interface (CLI) that allows users to perform targeted operations on the dataset without the need to download or analyze the entire dataset at once. The \dataset{} CLI is designed to be efficient and user-friendly, enabling users to retrieve and analyze specific subsets of GPT metadata based on their requirements.

\subsection{Data Retrieval} 
The CLI provides a command for retrieving GPT instances based on specific criteria using keyword search. Users can use the \texttt{python gptzoo.py -search} command followed by a set of keywords to query the dataset and retrieve matching instances. The command supports searching across various metadata fields such as name, author, category, tags, and more. The following example command demonstrates the search functionality:
\vspace{-2pt}
\begin{lstlisting}
python gptzoo.py -search --tags "programming" "software guidance" --description "software development"
\end{lstlisting}

The search results are saved in the ``results'' folder with a default timestamp filename.

\subsection{Data Analysis}
The CLI offers commands for performing analysis on specific subsets of the \dataset{} dataset. Users can use the \texttt{python gptzoo.py -analyze} command to calculate various metrics and statistics for GPT instances matching certain criteria. The command supports specifying a range of GPT instances using filters such as category, tags, rating, or custom keywords. The following example demonstrates the search functionality:
\vspace{-2pt}
\begin{lstlisting}
python gptzoo.py -analyze --tags "programming" "software guidance" --description "software development"
\end{lstlisting}

The analysis results, including the specified metrics, are displayed in tabular format, providing users with insights into the performance and characteristics of the selected GPT instances. Here's an example that showcases the analysis functionality:

\subsection{Command-Line Help}
The \dataset{} CLI provides comprehensive help documentation accessible through the \texttt{-help} flag. Users can append \texttt{-help} to any command to view detailed information about the command's usage, available options, and examples. The help documentation serves as a quick reference for users, assisting them in effectively utilizing the CLI's functionalities. The following example command demonstrates the help functionality:
\vspace{-2pt}
\begin{lstlisting}
python gptzoo.py -help
\end{lstlisting}

\section{Application Scenario}
\label{sec:scenario}

The \dataset{} dataset offers a unique foundation for researchers to explore the potential and impact of GPTs on various stakeholders, including GPT Store managers, developers, researchers, policymakers, and end-users. This dataset not only supports direct applications but also provides a rich context for broader research that benefits these groups indirectly.

\textbf{Insights for GPT Store managers.}
Research based on the \dataset{} dataset can uncover trends and patterns in GPT usage and preferences, which can inform GPT Store managers indirectly. Such research might reveal the features most valued by users, the performance metrics that correlate with user satisfaction, and gaps in the current market offerings. 

\textbf{Empowering developers through research.}
Researchers can analyze the \dataset{} dataset to identify best practices in GPT development and pinpoint which GPT features lead to successful applications across various domains. This research can be shared with developers, offering them insights that influence their design and development strategies. Such indirect benefits can accelerate innovation and improve the quality of customized GPT-powered applications, ensuring they meet the evolving needs of users.

\textbf{Advancing research and informing policy.}
The comprehensive metadata in the \dataset{} dataset enables researchers to study the evolution and effectiveness of GPTs, providing evidence-based insights that can inform both academic research and policy decisions. Policymakers can benefit from research that assesses the societal impact, ethical considerations, and regulatory implications of GPT technologies. This can help in crafting policies that promote beneficial AI uses while mitigating risks associated with privacy, security, and equity.

\textbf{Enhancing user experience.}
Research leveraging the \dataset{} dataset can foster the creation of enhanced tools and methodologies for evaluating and comparing GPTs. These advancements can indirectly benefit users by refining the selection tools they utilize to choose GPTs. Consequently, this research can improve decision-making support, enabling users to more effectively select GPTs that align precisely with their specific needs and tasks.

\section{Conclusion}
\label{sec:conclusion}
This paper introduces \dataset{} dataset, a comprehensive large-scale dataset comprising \sumofgpts{} GPT instances, each enriched with extensive metadata and core content, including instructions, knowledge files, and third-party services. This rich dataset is designed to facilitate detailed analysis and comparison of GPTs, providing a robust foundation for understanding their capabilities and applications. Additionally, we present an automated CLI tool engineered for efficient keyword-based retrieval and analysis of these GPT instances. This tool empowers researchers to delve deeper into the vast potential of GPTs, driving significant advancements in GPT-related research.

\section*{Ethics}
The \dataset{} dataset is exclusively available to researchers and only supports requests for research purposes. To access the dataset, please visit our repository at \url{https://github.com/security-pride/GPTZoo}. Content, including instructions, knowledge files, and third-party services associated with GPTs, is \textbf{temporarily unpublished} due to ethical concerns. Special requests will be considered on a case-by-case basis.

\balance

\bibliographystyle{plain}
\bibliography{main}

\end{document}